\documentstyle[12pt]{article}
\newfont{\feff}{cmti10}
\def\undertext#1{\vtop{\hbox{#1}\kern 1pt \hrule}}

\begin{document}

\begin{titlepage}
\title{ \begin{flushright} {\bf\normalsize
PUPT-1311}\\ \end{flushright} Critical Behavior of Dynamically
Triangulated Quantum Gravity in Four Dimensions}

\author{ M.E. Agishtein, A.A. Migdal\\ Physics Department and\\
Program in Applied and Computational Mathematics\\ Fine Hall,Princeton
University\\Princeton,NJ,08544-1000\\ } \date{March 1992}

\maketitle

\begin{abstract}
We performed detailed study of the phase transition region in Four
Dimensional Simplicial Quantum Gravity, using the dynamical
triangulation approach. The phase transition between the Gravity and
Antigravity phases turned out to be asymmetrical, so that we observed
the scaling laws only when the Newton constant approached the critical
value from perturbative side.  The curvature susceptibility diverges
with the scaling index $-.6$. The physical (i.e. measured with heavy
particle propagation) Hausdorff dimension of the manifolds, which is
2.3 in the Gravity phase and 4.6 in the Antigravity phase, turned out
to be 4 at the critical point, within the measurement accuracy.  These
facts indicate the existence of the continuum limit in Four
Dimensional Euclidean Quantum Gravity.
\end{abstract}
\end{titlepage}

\newpage
\renewcommand{\thesection}{\Roman{section}}

\section{Introduction}

The Quantum Gravity remains the last unsolved mystery of twentieth
century physics. Unlike, say, turbulence, where we at least know what
is happening and what are the equations to solve, in Quantum Gravity
we cannot be sure about the basic principles.

One would like to provide the meaning to the Euclidean
functional integral corresponding to the Einstein Action:
\begin{equation}
{\cal Z}(\lambda,G)=\int {\cal D}g\exp
\left( 	 -\int d^4 x\sqrt{g}
\left( 	\lambda -
\frac{R}{ G}
\right)
\right)
\label{ZLG}
\end{equation}
This integral formally diverges, since the curvature $R$ is not
bounded (neither from above nor from below). Moreover, there is no
unique way to define the continuous measure ${\cal D}g$ in infinite
dimensional configuration space of metric fields. The perturbative approach
does not lead to the renormalizable theory which means that the
genuine nonperturbative definition should be given before any
computations could be done.

    These problems, however, don't arise in Dynamical Triangulation
(DT) approach to Quantum Gravity (QG). The DT technique was introduced
several years ago in  two dimensional case (QG2) as a generalization of
Regge calculus~\cite{BKKM,BD,ADF}.  Metric field was represented by
simplicial complex built from equilateral simplexes, the integration
over $g$ being defined as summation with equal weight over all
simplicial complexes of given topology. In two dimensional case this
procedure would correspond to the summation over all possible
triangulations, which gave the name DT to
the method.

    In the original Regge-Ponzano approach, the lattice structure was
fixed (usually regular), and link lengths were the fluctuating
dynamical variables, subject to triangle inequalities. It does not
seem possible to define general covariant measure in Regge calculus,
in a sense, that there is no symmetry to take place of the
diffeomorphism group of the classical continuum theory.  Even though
it is much easier to implement than DT, and one can obtain some
exciting results with Regge Calculus~\cite{Hamber,Hamber2}, their
meaning is yet not clear. The big question is how do these results
depend upon the chosen lattice structure. There is no apparent reason
to believe in universality in case of broken general covariance.

    The DT approach is, however, generally covariant by construction,
with the simplex permutations as discrete analog of the
diffeomorphism.  When we sum over all connectivity matrices at given
topology, all simplexes are equivalent, hence there is permutational
symmetry. The labels, numbering simplexes, play the role of discrete
coordinates, so the relabeling of the same simplicial manifold
corresponds to diffeomorphism~\footnote{To be more precise, the diffeomorphism
represent the subset of all permutations, which becomes smooth transformations
in naive continuum limit.}.

 The local curvature of such system is
defined as the deficit angle of the loop of simplexes surrounding two
dimensional hinge, and it is bounded from above. The net curvature is
also bounded from below at fixed volume, as we shall discuss later.
Therefore, the functional integral is well defined, at least for fixed
volume and fixed topology. If there is a continuum limit in such a
theory, it will be a real physical limit.

    The first confirmation of the method came from the Liouville Field
Theory (LFT): the values of critical exponents computed analytically
in the DT model were obtained from purely continuous arguments.  The
early computer simulations were not accurate enough to verify these
values, but at least proved the concept. The basic elementary moves
found in \cite{BKKM} proved to be quite useful.

    The later investigations of the issues of internal and external
geometries in DT and its correspondence with LFT
predictions/assumptions~\cite{amigo2,radi} showed, that the DT model
in continuum limit indeed corresponds to QG2. At the same time some
inherent inconsistencies in the conventional view on the quantum
geometry based on classical geodesic lines were found\cite{David}.

    The remarkable advantages of DT over continuum methods in two
dimensions, caused the attempts to generalize it to higher
dimensions.  In three dimensions~(QG3), the dynamical triangulations
were simulated recently by three groups \cite{qg3,qg32,bjorn,bult}.
All three simulations show the existence of statistical limit, which
implies that the number of triangulations is exponentially bounded.
In \cite{bult} the first order phase transition was observed, with two
different values of average curvature, both finite in lattice units,
which implies the absence of continuum geometry. This transition was
later on investigated and confirmed on much larger systems
in~\cite{AVBK,AV}.  In \cite{qg3,qg32} the average curvature was
forced to continuum limit, by extra constraint, which corresponds to
unstable phase of \cite{bult}. The absence of continuum limit in QG3
is not so surprising, as there are no gravitons in 3
dimensions.\footnote{As for the topological gravity in 3 dimensions,
most likely it corresponds to Regge-Ponzano-Turayev-Viro model, where
the $ 6 j $ symbols attached to tetrahedra eliminate necessity of
summation over triangulations.  The model then reduces to finite
simplicial complex, say, 2 tetrahedra for the spherical topology, with
no continuum space. Apparently, we are dealing with different model.}
%

    In the four dimensional case the situation was different.  The
first simulations~\cite{AM4} revealed the existence of nontrivial
phase transition, which looked like a second order one, with small
hysteresis, the significance of which was not clear at that time. The
existence of the second order phase transition would imply
renormalizability and existence of continuum limit. These results were
later qualitatively confirmed on much smaller statistics by two other
groups~\cite{AJ4,V4}.

    In this work we addressed ourselves to detailed investigation of
thermodynamics and internal geometry of QG4 in the critical region. We
studied the four-spheres with up to $32 K $ simplexes and observed,
that the scaling laws, which characterize second order phase
transition, exist only when the system approaches critical point from
the perturbative phase (small gravitational constant and positive
curvatures). When we approached the critical point from the
nonperturbative phase of large negative curvatures, we did not find
convincing evidence for the powerlike singularities.  The physical
geometry, measured on this system also showed very interesting
behavior. The Hausdorff dimension $d_h$ of the QG4 manifolds in the
perturbative phase is about $d_h \approx 2.3$, which correspond to
``compactified'' space. In the nonperturbative phase, $d_h\approx
4.6$, while at the phase transition point it is $d_h=4$ within the
accuracy of the measurements.

    These results allow for the exciting physical interpretation: We
simulated the asymptotically Euclidean QG4 from the first principles.
Unlike 2 and 3 dimensional cases, we observed the existence of
nontrivial continuum behavior. When the system approaches the critical
point from the gravity side, the scaling properties of intrinsic quantum
geometry of the manifolds presumably become the ones of flat four-dimensional
space.

    The paper is organized as follows.	 In the Section II we describe
the kinematics of four-dimensional manifolds, basic steps of the Monte
Carlo simulations, and the main observables.  In the Section III we
present the results, which are followed by the Discussion in the
Section IV.

\section{Numerical method}

    The numerical technique of QG4 simulations is described in details
in~\cite{AM4}. For the reader's convenience we repeat the basic facts
below.

    We simulated the functional integration~(\ref{ZLG}) by a Markov
chain in the space of four dimensional simplicial manifolds. The
manifolds are represented by simplicial complexes built from
equilateral simplexes with unit volumes.  Each 4D simplex has 5
tetrahedra faces, 5 vertices, 10 links and 10 triangles.  Simplicial
complex is a collection of these simplexes with pairwise identified
faces. Simplicial manifold is a simplicial complex with the extra
condition: for any given vertex, the set of simplexes sharing this
vertex should form a spherical ball.

    The dynamical triangulation approach is a modified version of
Regge calculus~\cite{misner}. The space in considered flat inside the
$D=4$ simplexes, the curvature being concentrated in $D-2=2$
dimensional hinges, i.e. triangles.  The angle between two
tetrahedra-faces, sharing a triangle is
\begin{equation}
	\alpha= \arccos(\frac{1}{D})= 1.3181161
\end{equation}

    The volumes of $K$-simplexes are given by
\begin{equation}
	V_K= a^K \Omega_K=\Omega_K (\frac{V_D}{\Omega_D})^{\frac{K}{D}} ; \;
	\Omega_K=\sqrt{\frac{\Gamma(K+2)}{2^K\Gamma^3(K+1)}}
\end{equation}
We are going to measure volumes in the units of $V_4$, lengths -- in
$a=2*6^{\frac{1}{4}}*5^{-\frac{1}{8}} V_4^{\frac{1}{4}}$, and
curvature in $\frac{V_2}{V_4}=3*2^{\frac{1}{2}}*5^{-\frac{1}{4}}
V_4^{-\frac{1}{2}}$.  The local curvature, i.e. deficit of angle on a triangle
$\triangle$ which is shared by $ N_4(\triangle) $ 4-simplexes equals to
\begin{equation}
	R(\triangle) = 2 \pi - \alpha N_4(\triangle)
\label{}
\end{equation}
$\alpha$ not being rational part of $ \pi $ makes it impossible to construct a
locally flat space from equilateral simplexes. Still, if the volume of the
system, which is now equal to
the number of 4D simplexes in the manifold, is large enough, the
manifold can be flat in average.\footnote{Anyway, the observable metric and
curvature get renormalized due to strong quantum fluctuations.}
In such a complex, the average number of simplexes $<N_4(\triangle)>$ sharing a
triangle would be

\begin{equation}
    <N_4(\triangle)>=\frac{2\pi}{\alpha}=4.7667921
\end{equation}

    Let $N_k$ , $k=0,1,2,3,4$ be the net numbers of $k$-dimensional
simplexes in the manifold.  For $D=4$ the Euler relations read:
\begin{equation}
	5 N_4= 2 N_3
\label{E1}
\end{equation}
\begin{equation}
	\sum_{k=0}^{4} (-1)^{D-k} N_k =2
\label{E2}
\end{equation}

    The manifold condition provides additional relations between the
invariants $N_k$. These relations were derived in ~\cite{fein} for simplicial
manifolds of any dimension. We will
need only the simplest one:
\begin{equation}
	2N_1-3N_2+4N_3-5N_4 = 0
\label{E3}
\end{equation}
    This relation is independent from~(\ref{E1}),(\ref{E2}).  We,
therefore, can choose the volume $N_4$ and the number of vertices $N_0$ as our
parameters.  Using~(\ref{E1}),(\ref{E2}),(\ref{E3}) we
can express the integral  curvature of the manifold in terms of $N_0,N_4$ as
follows~\footnote{ As it was mentioned above, the conventional
normalization of curvature contains the extra factor
$3*2^{\frac{1}{2}}*5^{-\frac{1}{4}}\approx 2.84$, which we dropped.}:
\begin{equation}
	R = \sum_{\triangle} R(\triangle)= 2\pi N_2 - 10\alpha N_4=
	4\pi (N_0+N_4-2) - 10\alpha N_4
\label{E4}
\end{equation}

    We simulated the functional integration~(\ref{ZLG}) by Markov
chain in the space of QG4 simplicial manifolds.  As was mentioned
in~\cite{qg3} the minimal set of elementary moves, which guarantees
for this chain to be ergodic can be obtained from all possible
decompositions of the boundary of $D+1$ simplex into two $D$-complexes.
At $D>4$ this is a trivial corollary of Smale's $h$-cobordism theorem.  It
means,
that there are $D+1$ basic moves in $D$-dimensional gravity. For $D=4$
we brought them together in the table~\ref{mv4}, where we, for the
sake of simplicity, represented a simplex by the list of its vertices in
the alphabetical order.  These elementary moves are theoretically
ergodic and allow to construct a Markov chain in the space of
simplicial complexes.  The moves described in the Table~\ref{mv4} can
sometimes lead to degenerate triangulations, which are  excluded from
conventional definitions of simplicial manifold. For example, one could get
several disconnected sets of 4-simplexes, sharing the same triangle, or some
other simplex of dimension $ k < 4 $. So, like in QG2 and in QG3
simulations, we had to check the local geometry at each step and
reject the improper moves. Let us stress, however, that the manifold
restriction is not absolutely necessary from the continuum limit point of view,
like in QG2 one
could allow tadpoles and self-energy graphs. We eliminated degenerate
configurations because it simplified the representation of triangulation in
computer memory, and also because we expected the degenerate triangulations to
delay the infinite volume limit.
\begin{table}
\begin{tabular}{||c|c|c||c|c|c|c|c||} \hline
\multicolumn{8}{||c||}{\bf QG4}\\ \hline
move & smplxs before  & smplxs after & $\Delta N_0$ & $\Delta
N_1$ & $\Delta N_2$& $\Delta N_3$ & $\Delta N_4$ \\ \hline
bar. subd. & ABCDE & ABCDF, BCDEF, & 1 & 5 & 10& 10 & 4\\
    &       & ABDEF, ABCEF     &   &   &  &  & \\
    &       & ACDEF     &   &   &  &  & \\ \hline
bar. subd. & ABCDF, BCDEF, & ABCDE & -1 & -5 & -10& -10 & -4\\
removal & ABDEF, ABCEF  &      &   &   &  &  & \\
    & ACDEF  &         &   &   &  &  & \\ \hline
flip & ABCDE, ABCDF, & BCDEF, ACDEF, & 0 & 0 & 0& 0 & 0\\
    & ABCEF & ABDEF &          &   &   &  &   \\ \hline
two-four & ABCDE, BCDEF & ABCDF, ABCEF, & 0 & 1 & 4& 5 & 2\\
exchange &   & ACDEF, ABDEF    &   &   &  &  & \\ \hline
four-two & ABCDF, ABCEF, &ABCDE, BCDEF &  0 & -1 & -4& -5 & -2\\
exchange &    ACDEF, ABDEF    & &   &   &  &  & \\ \hline
\end{tabular}
\caption{Ergodic moves in QG4}
\label{mv4}
\end{table}

    We used the Grand Canonical algorithm proposed in~\cite{qg32},
which is a modification of the algorithm introduced in~\cite{bjorn}.
The Action~(\ref{ZLG}) was changed  as follows:
\begin{equation}
    {\cal Z}(\hat{\lambda}_4,\hat{\lambda}_0) =
    \sum_{manifolds} \exp \left(
    -\hat{\lambda}_4 N_4 - \hat{\lambda}_0 R
    -\Delta\lambda_4 (N_4-\hat{N}_4)^2
    -\Delta\lambda_0 (R-\hat{R})^2
    \right)  \label{Dlam}
    \end{equation}
$$
= \sum_{N_4,N_0} {\cal Z}(N_4,N_0)
    \exp \left( -\hat{\lambda}_4 N_4 - \hat{\lambda}_0 R
    -\Delta\lambda_4 (N_4-\hat{N}_4)^2 -\Delta\lambda_0 (R-\hat{R})^2
    \right) \
$$
where $N_4$ is the volume of the system, $R$ -- the net curvature,
and ${\cal Z}(N_4,N_0)$ -- the microcanonical partition function, which is
nothing but the number of the corresponding triangulations of the 4-sphere. The
$\hat{N}_4$ and $\hat{R}$ are the desired volume and the net curvature
respectively.  This modification allows to take some guess values of
$\lambda_i$ and to measure the functions $\lambda_i(N_4,R),i=0,4$
directly from the saddle-point equations.
\begin{equation}
    \frac{\partial \ln({\cal Z}(N_4,N_0))}
    {\partial N_4} \equiv
    \lambda_4(N_4,R)= \hat{\lambda}_4+2\Delta\lambda_4 (N_4-\hat{N}_4)
    \label{srul4}
\end{equation}
\begin{equation}
    \frac{\partial \ln({\cal Z}(N_4,N_0))}
    {\partial R} \equiv
    \lambda_0(N_4,R)= \hat{\lambda}_0+2\Delta\lambda_0 (R-\hat{R})
    \label{srul0}
\end{equation}
Varying $\Delta\lambda_i$ one can smoothly interpolate between
Grand Canonical and Micro Canonical simulations. From one hand, we
would like the system to perform large steps in the phase space, i.e.
to allow $N_4$ and $N_0$ to vary in wide enough limits. From the other
hand, we want to be far from the finite size effects and to collect
sufficient statistics near $\hat{N_i}$, where the saddle point
approximations~(\ref{srul4}, \ref{srul0}) are valid. For these
technical reasons we used $\Delta\lambda_4=\Delta\lambda_0= 0.005$.

One could regard these equations as parametric relations between
average values of $ N_4, R $ and desired values $ \hat{N}_4, \hat{R}
$. Taking different values of $ \hat{N}_4, \hat{R} $ we could measure
average values and use above relations to find the effective values $
\lambda_4(N_4,R), \lambda_0(N_4,R)$. The results should be independent
of auxiliary parameters $ \hat{\lambda}_0,
\hat{\lambda}_4,\Delta\lambda_4,\Delta\lambda_0$ as long as we stay in the
region of large volumes and curvatures and small fluctuations, so that the
saddle point equations apply. The disadvantage of
this method would be the
necessity to interpolate functions of two variables  $ N_4, R $ given
at scattered points. We preferred to fine tune  $ \hat{N}_4, \hat{R} $
so that they coincide with average values. Then we could plot fixed
volume curves in the $  \hat{\lambda}_0,\hat{\lambda}_4 $ plane, and
observe the convergence to the infinite volume limit.

We performed our simulations in the following way: $\hat{\lambda}_0$ was
slowly varied from large negative values (perturbative phase) to small
positive values (nonperturbative phase) and back, keeping $N_4$ fixed.
This way, we simulated each value of $\lambda_0$ twice, coming from
different directions.  If there were a first order phase transition,
or we did not thermalize the system at given $\lambda_0$ we would
observe hysteresis.  We observed some hysteresis in the
beginning~\cite{AM4}, but we were not sure, what was the reason for
it.  By performing longer runs, we succeeded in eliminating it, which
became our thermalization criteria. Since we were interested in the
critical point vicinity, we had to take a fine grid in $\lambda_0$
(usually 60 values)~\footnote{ We could, of course, simulate very few
points and perform Ferrenberg-Swendsen interpolation like
in~\cite{V4}. This method, however, uses the assumption of regular
second order phase transition and its accuracy is questionable.  We
decided, therefore, to employ a straightforward approach, which
allowed us to observe the asymmetry of the phase transition.} In order
to make the error bars in the measurements of $<R(\lambda_0)>$ and
$<\lambda(\lambda_0)>$ smaller than the variation of these observables
between the neighboring values of $\lambda_0)$ we had to perform
160000 sweeps at each value of $\lambda_0)$.

\section{Results}

There were two major aspects, we were
concerned about in our simulations: thermodynamical properties of QG4
near the phase transition and physical geometry at the critical point.

Using the formulae~(\ref{srul4}),~(\ref{srul0}) we obtained the curves
for $<R(\lambda_0)>$ and $\lambda(\lambda_0)$ for different values of
volume $N_4$ (see Fig.~\ref{raver}, ~\ref{lambda}).

These plots clearly demonstrate the difference between Gravity and
Antigravity phases. In the Gravity phase
$(\lambda_0 < \lambda_0^{critical})$ the curves converge to some
infinite volume limit.  At large negative $\lambda_0$ this phase
apparently corresponds to perturbative continuum gravity.  In the
Antigravity phase, however, $-<R(\lambda_0)>$ grows with volume, being
bounded only by the lattice finite size effects.  The physical
behavior, therefore, exists only on the Gravity side of the phase
transition, the Antigravity side corresponding to the branched polymer
phase and formation of baby-universes.

Susceptibility measurements (see Figure~\ref{fit_r}) showed very
clearly, that the phase transition is not symmetric.
Since the phase transition did not look like the traditional second
order transition, we did not want to make our conclusions based on
these noisy data or on
extrapolation from 5-6 points. We collected large statistics and fitted
$\lambda_0$ with the fifth degree polynomial $\lambda_0\approx
P_5(<R>)$, like in Mean Field theory.  Susceptibility $\chi$ is
calculated from the derivative of $P_5$ as follows:
\begin{equation}
 \chi=-\frac{d<R>}{d\lambda_0}=-\frac{1}{P_5^\prime(<R>(\lambda_0))}
\end{equation}
The critical value of $\lambda_0^c$ is obtained from the (real part of
the nearest to the real axis) zero of polynomial $P_5^\prime$ and
corresponds to the maximum of $\chi$.  We illustrated this procedure
on Figure~\ref{fit_r}.

We performed the finite size scaling analysis of $\chi^c$, fitting it
with the law $a+b N_4^\mu$. The fit showed that $\chi^c$ grows as
small power of $N_4$, $\mu\approx 0.14$, which is consistent with the
logarithmic growth.  The fit of $\lambda_0^c$ with $a+b N_4^\nu$was
consistent with the hypothesis $\nu\approx -0.5$ (see
Figure~\ref{finsiz}).

The  behavior of $\chi$ near the singularity is described by
the critical index $\gamma$:
\begin{equation}
	(\chi^c-\chi) \sim |\lambda_0^c-\lambda_0|^{-\gamma}
\end{equation}
This observable had rather big errors in our simulations, this is why we
decided  to calculate  $\gamma$ directly from the $<R>$ measurements:
\begin{equation}
	|R^c-R| \sim |\lambda_0^c-\lambda_0|^{1-\gamma}
\end{equation}
This observable was the most interesting one.  The phase transition
turned out to be asymmetric, with unequal $\gamma_<$ and $\gamma_>$ on
the different sides of $\lambda_0^c$ (see Figure~\ref{two_lam}). The
measurements of $\gamma$ for different system sizes are brought
together in the table~\ref{gamma}.
\begin{table}
\begin{center}
\begin{tabular}{||c|c|c||} \hline
$N_4$ & $\gamma_>$ & $\gamma_<$ \\ \hline
1024 & 0.184853 & 0.368676\\
2048 & 0.142918 & 0.458431\\
3076 & 0.151671 & 0.431315\\
4096 & 0.150384 & 0.516641\\
6144 & 0.116637 & 0.499929\\
8192 & 0.156094 & 0.548239\\
12000 & 0.070800 & 0.543294\\ \hline
$\infty$ & $\approx 0$ & $\approx .6$\\ \hline
\end{tabular}
\caption{Measurements of $\gamma$ for different system sizes and
extrapolation to the infinite volume.}
\label{gamma}
\end{center}
\end{table}

The weak singularity in the Antigravity phase $\gamma_>\approx .1$,
is consistent with the visual observation, that the lines of
average curvature are almost straight lines to the right of the
inflection point.

After separating the phases and investigating the thermodynamics
of the phase transition, we can ask a natural question: what are
these phases all about? In the spin models, hot phase means
completely random orientation of neighboring  spins, while
the cold phase is dominated by large domains. What can be the
order parameter in the pure gravity?  There is nothing, but
geometry fluctuations in this theory, so can geometry itself be
the order parameter?

As we already mentioned in the first paper~\cite{AM4}, the
internal Hausdorff dimension $d_h$, measured in physical definition
of distances, has completely different value in both phases and
can be considered as an order parameter for QG4, in a sense that in
the Gravity phase the space tents to compactify, so that $d_h <4$
whereas in the Antigravity phase the space tends to branch like the
tree, so that $d_h>4$. If the DT model describes continuum gravity,
then we must have $d_h=4$ at the transition point.

Therefore, the local parameter
\begin{equation}
	\eta = \lim_{l \rightarrow 0} \frac{V(l)}{l^4}
\label{}
\end{equation}
vanishes in the Antigravity phase, goes to $ \infty $ in the Gravity phase, and
remains finite at the critical point. This is the closest we could get to the
order parameter.

To define the distance we consider the propagation of massive
test particle as first suggested by~\cite{David}.
When the propagator is exponentially small, it decays as
\begin{equation}
  G_{1 2} \rightarrow A \left(l_{12} \right)\exp
	\left(
	  - m l_{12}
	\right)
\end{equation}
where $ l_{12} $ is the geodesic distance between two points.
The factor $ A $ is not universal, but the mass m is.
This is
the physical mass of the particle, with the gravitational
renormalization. This renormalization is {\bf multiplicative} rather
then additive, as it follows from the fact that discrete Action
\begin{equation}
  \sum_{<i j>} \left(\phi_i-\phi_j \right)^2
	+\sum_i m_0^2 \phi_i^2 = \sum_{ij} \left(
	\left(m_0^2 + D+1 \right) \delta_{ij} - S_{ij}
	\right) \phi_i \phi_j
\label{partact}
\end{equation}
where $ S_{i j} $ is the simplexes connectivity matrix, gets additional
translational symmetry
\begin{equation}
  \phi_i \rightarrow \phi_i + const
\end{equation}
at vanishing bare mass $ m_0 $. In other words, at vanishing bare mass
there is always the zero mode $ \phi_i = const $, so that the physical
mass would also vanish. In this respect the gravity is simpler
then the usual gauge theory where there was additive renormalization
of the scalar particle mass.

In terms of the random walk the propagator
\begin{equation}
	G_{1 2} = \left(\left(m_0^2 + D+1 \right) \hat{I}  - \hat{S}\right)^{-1}_{1 2}
	= \frac{1}{D+1+ m_0^2}\sum_{k=1}^{\infty}
	\left(\frac{\hat{S}}{D+1 + m_0^2}\right)^k_{1 2}
	\label{}
\end{equation}
corresponds to the process of hopping from simplex to one of its $ D+1 $
neighbors with probability $ ( D+1+ m_0^2)^{-1}$. The net hopping probability
\begin{equation}
	\frac{D+1}{D+1+ m_0^2} \equiv \exp (-m )
\label{}
\end{equation}
defines the particle mass in the sum over paths
\begin{equation}
	G_{1 2} = \sum_{\Gamma_{1 2}} \exp\left(- m l\left[\Gamma_{1 2} \right]
\right)
\label{}
\end{equation}

One could use two heavy propagators with masses $ m,m'$ to define the geodesic
distance
\begin{equation}
  l_{12} =  \frac{\left(\ln G_{1 2}- \ln G_{1 2}^{'}\right) }{m'-m}
\end{equation}

If one takes the local limit after the large mass limit, he will
obtain the ordinary lattice geodesics. This metric, which we called
``mathematical'', does not have a continuum limit: it develops
parametrically large amount of singularities, $d_h$ grows with the
test-sphere size etc.~\cite{amigo,amigo2,qg3}.

However, if the local limit is taken before the large mass limit, then
physical geometry differs from mathematical one. The entropy of the
paths leads to effective transverse thickness of the dominant set of
paths. This thickness is not related to the cutoff, and therefore the
thick path is not sensitive to the lattice artifacts, such as the
embryonic Universes, observed in the mathematical geometry.

We used the ``Black-Box'' Multigrid method to solve the Poisson
equation on random manifolds (see \cite{AM4} for details) and obtained
the volume-radius histogram for the manifolds near the critical point
(Figure~\ref{haus_cr}). We fitted them with the scaling law $<V(r)>
\approx a+br^{d_h}$.  The result was quite remarkable: as we already
mentioned above, $d_h=2.3$ far in the Gravity phase and $d_h=4.6$ in
the Antigravity phase, but at the phase transition we found $d_h = 4$,
just as expected!

%
\section{Discussion}

For somebody with the Classical Relativity background our model must look like
a mathematical toy, nothing to do with real 4D gravity. All we did, we
generated
random collection of simplexes with spherical topology and fixed number of
vertices. How could that be the same, as "sum over all graviton loops" of
continuum theory?

Well, this miracle already happened in QG2 where random triangulation was
proven
to reproduce the " sum over all Liouville loops". The general covariant theory
looks much simpler without gauge fixing. As for effective infrared Lagrangean,
such as phenomenological  Einstein Gravity, it cannot be built in the Quantum
Field Theory, but rather should come out dynamically.

Let us now speculate about the physical meaning of our results. At this
point there are several possibilities. First of all, it is still
possible that there is no critical behavior. One have to simulate
much larger systems and check the trends we observed at the relatively
small systems of several thousand simplexes. In particular, one have
to check the observed asymmetry of the critical point.  Would the
average curvature continue to go down to $ - \infty $ in the
Antigravity phase? Would the scaling law for the curvature in the
Gravity phase remain the same for large systems?

Also, at larger systems we should be able to observe the running
Newton constant: would it grow or decrease with scale? In other words,
is our fixed point infrared or ultraviolet stable? If the $ 2 +
\epsilon $ approach~\cite{KN} could be taken as a
guide to the QG4, then there must be the ultraviolet stable fixed point,
corresponding to the conformal field theory, and the infrared stable
fixed point corresponding to zero gravity.
At our modest scales all we could see, is the ultraviolet one, but
with one more decade of scales one should see the decrease of
effective Newton constant.

With alternative scenario of asymptotically
free Gravity with $ R^2 $ terms one would observe the growth of
effective Newton constant. This scenario seems unlikely though, as we
did not introduce large $ R^2$ terms in the bare Action, as we should
have done in the asymptotically free theory. Clearly, we are studying
the strong gravity at small distances, with zero or at least small $
R^2$ terms in effective Action. The apparent coexistence of continuum
space with $ d_h=4 $ at large scales with nontrivial thermodynamical scaling
indices fits the first scenario better.

Some authors~\cite{AJ4,V4} claim to obtain direct correspondence
between the results of DT and Regge calculus~\cite{Hamber,Hamber2}.
We would like to emphasize, that we don't see any similarity.  And not
only the numbers are different ($\gamma_>\approx0.4$ in \cite{Hamber2},
$\gamma_<\approx0.6, \gamma_>\approx 0$ in our simulations). There is
no weak coupling phase in the Regge calculus, but this was the only
physical phase we observed in our simulations! The phase, where the
simulations of \cite{Hamber2} were performed does not have a continuum
limit in our model, and there is no significant singularity when one
approaches the critical point from this side.

The observed scaling laws in our model imply, that there are some
effective massless fields at the critical point. The natural
candidates are gravitons, but we have not proved this yet. All we
know, are the scaling laws, but not the structure of the operator
algebra. The graviton has spin 2, but we have first to find the global
$ O(4) $ rotational symmetry to be able to talk about spin.

The problem is that initially we do not have any fields in our theory,
but rather pointers from each simplex to its neighbors. These pointers
are our only degrees of freedom, which makes theory easy to implement
on the computer, but hard to interpret analytically. We need the reper
fields and the spin connections to define the $ O(4) $ group.

There is the natural local reper in any simplicial complex, namely,
each simplex has $ D+1=5 $ unit vectors $E_{\mu}^{(A)},A=1,\dots,5$
normal to its faces. They are linearly dependent, namely, they sum up
to zero in our case of equilateral simplexes. Each of above-mentioned
pointers is associated with corresponding normal vector. One could
redefine the pointers without changing the simplicial complex, namely,
one could rotate each simplex around its center, and then reconnect
with neighbors. These local rotations form the discrete group, which
is our counterpart of the local $ O(4) $ symmetry. As usual in the
lattice theory, we hope that in continuum limit only quadratic Casimir
invariant of this group would remain relevant, which would raise the
symmetry up to the full $ O(4) $ group, at least globally.

We could now add matter to our system, by placing the matter fields in
the middles of simplexes and coupling them to the neighbors in an
obvious way, using this reper. The spin connections $ \Omega_{<ij>} $
belonging to the corresponding representation of the simplex rotation
group, should be associated with each pair $ <ij> $ of neighbors, so
that local rotations of the reper would result in the gauge
transformations of the spin connections.

There are various alternative treatments of the reper and spin
connections. One could classically relate them to the metric, say, by
taking proper initial values at the beginning of the creation of the
simplicial manifold, when it reduced to the pair of simplexes (for the
spherical topology), and then rotating them correspondingly at each of
above local canonical moves. Alternatively, one could treat them as
independent dynamical variables, and simulate them along with the
triangulation. In presence of dynamical spinning matter, these
variables would interact with triangulation.

There are many interesting things to do with this model, like studying
Wilson loops corresponding to the spin connections, or the physical
Newton constant defined as susceptibility with respect to mass of the
test particle. Or one could study physical geodesic lines formed by
paths of heavy particles, and measure sum of  angles of geodesic
triangles. These studies would hopefully bring us closer to real
physical questions, which we do not know how to formulate so far: what is
unitarity, and how to come back to Minkowskian space.

    And the last point. In QG2 the basic moves were proven to be
ergodic. There is also a proof~\cite{BKKM}, that any two
configurations of the same genus can be connected by the number of
moves linearly bounded by the volume. It means, that we have finite
probability to reach any configuration of given volume in simulations.
In the higher dimensions the situation is more complicated. There is
little doubt, that the basic moves are ergodic, even though there is
no $h$-cobordism theorem in four dimensions. The problem of complexity
is, however, very much unclear. It has been proven~\cite{Nabut}, as a
corollary of the Novikov theorem of unrecognilizability of a sphere in
five and higher dimensions, that the number of moves, required to
transform one configuration into another is not bounded by any
computable function of the configuration volume. It means, that the
problem is ergodic in infinite time but it is infinitely complex.
There is no rigorous proof of this conjecture in four dimensions, but
it seems to be the same. This theorem, however, does not provide any
idea about the measure of the fraction of configurations, which can
not be reached in a functional time, so it can as well be zero.
Besides, we are interested only in the configurations, which are in a
sense close to the perturbative limit, so we might as well say, that
we are not interested in these weird branches. The problem, however,
is still open.

\section{Acknowledgements}

We are grateful to A. Ashtekar, D.Gross, S.Novikov, A. Peres, L.Smolin,
A.Voronov and E.Witten  for fruitful discussions and important remarks.  MEA
was supported by the Defense Advanced Research Project Agency under contract
N00014-86-K-0759. AAM was  partially supported by the National Science
Foundation under contract PHYS-90-21984.

    \newpage

    \begin{figure}[h]

    \caption{
    $<R(\lambda_0)>$ for $N_4=1024-12000$.
    }
    \label{raver}
    \end{figure}


    \newpage
    \begin{figure}[h]

    \caption{
    $<\lambda(\lambda_0)>$ for $N_4=1024-12000$.
    }
    \label{lambda}
    \end{figure}

    \newpage
    \begin{figure}[h]

    \caption{
    Direct measurements of susceptibility $\chi$.
    Fit of $\lambda_0$ with the polynomial of $<R>$ and susceptibility
    curve obtained from this fit.
    }
    \label{fit_r}
    \end{figure}

    \newpage
    \begin{figure}[h]

    \caption{
    Finite size scaling analysis  for $\chi^c$ and  $\lambda_0^c$.
    }
    \label{finsiz}
    \end{figure}

    \newpage
    \begin{figure}[h]

    \caption{
    Fit of the critical exponent $\gamma$:
	$\log|R^c-R| \approx |\lambda_0^c-\lambda_0|*(1-\gamma)$.
	The indexes on the different sides of the phase transition
    are clearly different.
    }
    \label{two_lam}
    \end{figure}


    \newpage
    \begin{figure}[h]

    \caption{ Geometry at the critical point:
    Hausdorff histogram: $\log - \log$ plot of the average
    volume inside a geodesic sphere versus its radius.
    Fit of the Hausdorff histogram by the power law:
    $<V(r)> \sim r^4$.  }
    \label{haus_cr}
    \end{figure}
    \newpage


\begin{thebibliography}{99}
    \bibitem{BKKM} D.V. Boulatov, V.A.Kazakov, I.K. Kostov, A.A.
    Migdal, Nuclear Physics {\bf B275} [FS17] (1986), 641.
    \bibitem{BD} A.Billoire, F.David, Nuclear Physics {\bf B275}
    [FS17] (1986), 617.
    \bibitem{ADF} J.Ambjorn, B.Durhuus, J.Fr\''{o}chlich,Nuclear
    Physics {\bf B270} [FS16] (1986), 457.
    \bibitem{Hamber} H.W.Hamber, Nuclear Physics {\bf B99A} (proc.  suppl.)
(1991).
    \bibitem{Hamber2} H.W.Hamber, Preprint UCI-90-60, November 1990.
    %
    \bibitem{amigo} M.Agishtein, A.A.Migdal, International Journal of Modern
Physics C {\bf 1}, No.1, April (1990), 165.
    \bibitem{amigo1} M.Agishtein,  L.~Jacobs,  A.A.Migdal, J.~Richardson, Mod.
Phys. Lett. A, {\bf 5}, 12, (1990).

    \bibitem{amigo2} M.E. Agishtein,  A.A. Migdal, Nuclear Physics B350 (1991),
690-728.
    \bibitem{David} F.David, Preprint RU-91-25.

    \bibitem{qg3} M.E.Agishtein, A.A. Migdal,  Mod. Phys. Lett. A Vol.~6, No~20
(1991) pp. 1863-1884, see also Errata.
    \bibitem{qg32} M.Agishtein, A.A.Migdal,  Numerical simulations
    of three dimensional quantum gravity, Nuclear. Phys. B,
    Proc. Suppl., 25A (1992), pp. 1-7.
    \bibitem{bjorn} J. Ambj{\o}rn, S.Varsted, Niels Bohr Inst. Preprint
NBI-HE-91-17.
    \bibitem{bult} D.V. Boulatov, A. Krzywicki, preprint LPTHE Orsay
    91/35.
    \bibitem{AV} J. Ambj{\o}rn, S.Varsted, Niels Bohr Inst. Preprint
NBI-HE-91-45.
    \bibitem{AVBK} J. Ambj{\o}rn, D.V. Boulatov,  A. Krzywicki,
    S.Varsted, Niels Bohr Inst. Preprint NBI-HE-91-46.
    \bibitem{AM4}M.E.Agishtein, A.A.Migdal,  Simulations of
    four dimensional quantum gravity. PUPT-1287. October 1991.
    \bibitem{KN} H. Kawai, M. Ninomiya, Nuclear Physics B336,
    pp. 115-145, (1990).
    \bibitem{AJ4} J. Ambjorn, J. Jurkiewicz, Preprint NBI-HE-91-47.
    \bibitem{V4} S.Varsted, Preprint UCSD/PTH 92/03, January-92.


    \bibitem{misner} Misner C.W., Thorne K.S., Wheeler J.A., Gravitation,
    Freeman and Company, New York,1973.
    \bibitem{fein} N.H.Christ, R.Feinberg, T.D.Lee, Nuclear Physics
    B202, pp. 89-125, (1982).
    \bibitem{KazBul} D.V.Boulatov, V.A. Kazakov, Phys. Lett., {\bf
184b},(1987), pp. 247-252.
    \bibitem{KazBul1} D.V.Boulatov, V.A. Kazakov, Preprint
    NBI-HE-88-42, (1988).
    \bibitem{radi} M.Agishtein, R.BenAv, A.A.Migdal, S.Solomon,
	Mod. Phys. Lett. A Vol.~6, No.~12, (1991), pp 1115-1132.
    \bibitem{ABMS} M.Agishtein, R.BenAv, A.A.Migdal, S.Solomon,
    Mod. Phys. Lett. A Vol.~6, No.~12, (1991),pp 1115-1132.
    \bibitem{mgross} M.Gross, S.Varsted, preprint NBI-HE-91-33.
    \bibitem{tobe} M.E. Agishtein,  A.A. Migdal, in preparation.
    \bibitem{Nabut} R. BenAv, A. Nabutovsky, private communication.
    \end{thebibliography}
\end{document}